\documentclass[pra,preprint]{revtex4}
\usepackage{epsfig}
\usepackage{amsmath}
\usepackage{epsfig}

\begin{document}
\title
{Solvable model of bound states in the continuum (BIC) in one dimension}
	\author{Zafar Ahmed$^{1,2}$, Sachin Kumar$^3$, Dona Ghosh$^4$, and Tarit Goswami$^5$}
	\affiliation{$^1$Nuclear Physics Division, Bhabha Atomic Research Centre, Mumbai 400 085, India \\ $~^2$Homi Bhabha National Institute, Mumbai 400 094,India	\\
		 $^3$Theoretical Physics Section, Bhabha Atomic Research Centre, Mumbai 400 085, India\\
		$^4$Department of Mathematics, Jadavpur University, Kolkata 700032, India\\
		$^5$Department of Computer Science and Engineering, Jalpaiguri Govt. Engineering College, Jalpaiguri, WB 735102, India
		}
	\email{1:zahmed@barc.gov.in, 2:sachinv.barc.gov.in, 3:rimidonaghosh@gmail.com, 4:taritgoswami456@gmail.com}
\date{\today}
\begin{abstract}
Historically, most of the quantum mechanical results have originated in one dimensional model potentials. However, Von-Neumann's Bound states in the Continuum (BIC) originated in specially constructed, three dimensional,  oscillatory, central potentials. One dimensional version of BIC has long been attempted, where only quasi-exactly-solvable models have succeeded but not without instigating degeneracy in one dimension. Here, we present an exactly solvable bottomless exponential potential barrier $V(x)=-V_0[\exp(2|x|/a)-1]$ which for $E<V_0$ has a continuum of  non-square-integrable, definite-parity, degenerate states. In this continuum, we show a surprising presence of discrete energy, square-integrable, definite-parity, non-degenerate states. For $E>V_0$, there is again a continuum of complex scattering solutions $\psi(x)$ whose real and imaginary parts though solutions of Schr{\"o}dinger equation yet their parities cannot be ascertained as $C\psi(x)$  is also a solution where $C$ is an arbitrary complex non-real number. 
\end{abstract}
\maketitle
In three dimensions, specially constructed  spatially oscillating central potentials are known to possess discrete energy Bound states in the Continuum (BIC) which
are called Von-Neumann states [1]. Though it started as a mathematical curiosity, later [2] these states have been realized in physical systems.

BIC have been studied mainly in optics [3]. In a photonic system consisting of two parallel dielectric gratings and two arrays of thin parallel dielectric cylinders [4], It has been shown that the interaction between trapped electromagnetic modes can lead to BIC. Such photonic systems find many applications in optics, such as amplification of electromagnetic fields within the photonic structures. Recently, the simultaneous occurrence of two fundamentally different phenomena, the BIC and the spectral singularity (time-reversed spectral singularity) has been achieved in a rectangular core dielectric waveguide with an embedded absorbing layer [5]. Two distinct types of optical BIC existing in broken and unbroken PT phase have also been classified in PT-symmetric optical waveguides [6].  A recent review article describing the possible experimental realizations and existing applications of BIC is also worth mentioning [7]. BIC can exist for electrons in solids. Electronic BIC has been proposed and studied in Condensed matter physics, They appear as points/lines in the momentum space and protected by the topological invariants [8,9]. The existence of BIC has been theoretically demonstrated as well in a quantum-dot system, in which BIC appear due to Fabry-Perot interference between the quasibound states of each dot [10], further an encryption device based on BIC has been theoretically investigated  in a topological Kitaev chain connected to a double quantum-dot structure [11].  In Ref. [12] authors have studied the BIC in hermitian tight bonding models, and shown that the transformation of a scattering state into a BIC can be formally described as a “phase transition” with a divergent generalized response function. A new type of Quantum phase transition triggering the magnetic BIC has been proposed [13] in graphene. The interplay between spin-orbit coupling and Zeeman splitting in atomic systems leading to the existence of BIC has been shown [14].  In Ref. [15] authors have analyzed strong coupling between excitons in transition metal dichalcogenides and optical BIC. Analytic solution of BIC for photons and atoms in a one-dimensional coupled-cavity array has been presented in [16]. Most recently, dynamic control of a BIC has been demonstrated in the dielectric meta-surfaces [17].

The link between the BIC and the Fano anomaly [18] in the
transmission spectra is another interesting phenomenon which has been investigated very
well [19].

For a long time one-dimensional version of BIC has been tried and the present work addresses this issue mainly. Very interestingly, in single channel, time-reversely symmetric quasi-one dimensional systems, surprising transmission-zeros have been observed earlier [19]. Here, we will have a new opportunity to see if this can also happen in a purely one dimensional system also.

Earlier, an interesting symmetric stair-case like potential  called pyramid barrier [20]  has been purported to be giving rise to BIC. However, a usual study of scattering from such a potential reveals that the pyramid potential cannot act any different from the rectangular barrier as both of them are bounded from below. By making these potentials very high or by making them large negative on both  sides, the uniform amplitude of asymptotic   oscillations in the scattering state can be made very small but not zero. So all the states in these two simple potentials remain scattering states which are  non-square-integrable. For higher order solvable models [21] of bottomless potentials like parabolic $V(x)=-x^2$ and triangular $V(x)= -|x|$ barriers, scattering solutions of Schr{\"o}dinger equation do vanish asymptotically yet they are not square-integrable.  For instance for the parabolic barrier $\psi(x) \sim e^{\pm ix^2}/\sqrt{|x|}$ [21].

Later, investigations  on Quasi-Exactly-Solvable (QES) 1D potentials have revealed [22-25] surprising results that  two square integrable states  can exist at one real energy $E=E_*$ in or above a bottomless barrier $(V(\pm \infty)= -\infty)$. This means a bound state exists even without two real turning points or it exists in a double barrier with four real turning points at $E=E_*$ where one usually expects [26] a complex energy resonance to occur. They even suggest a degeneracy at $E=E_*$ in one dimension. 
\begin{figure}[ht]
	\centering
	\includegraphics[width=4.5cm,height=5 cm,scale=1.3]{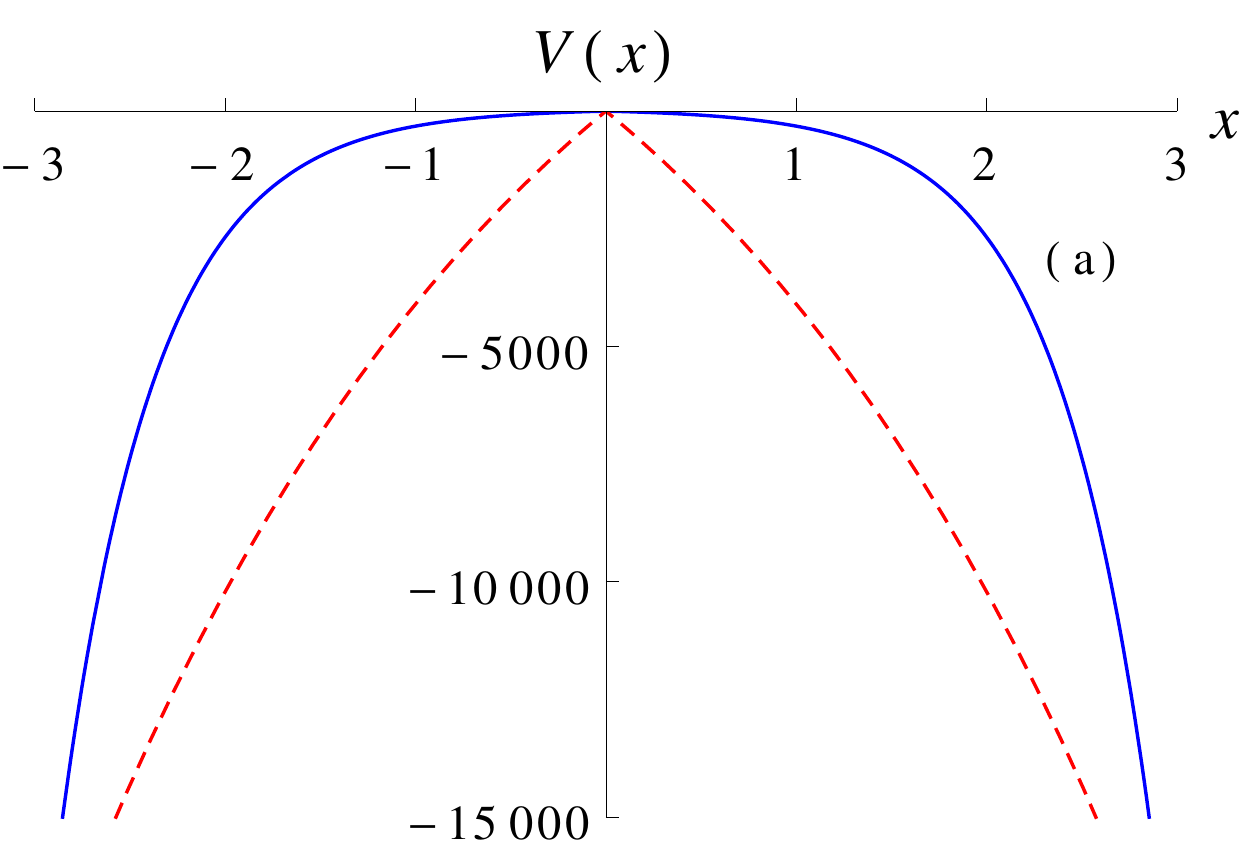}
	\hspace{0.5cm}
	\includegraphics[width=4.5cm,height=5 cm,scale=1.3]{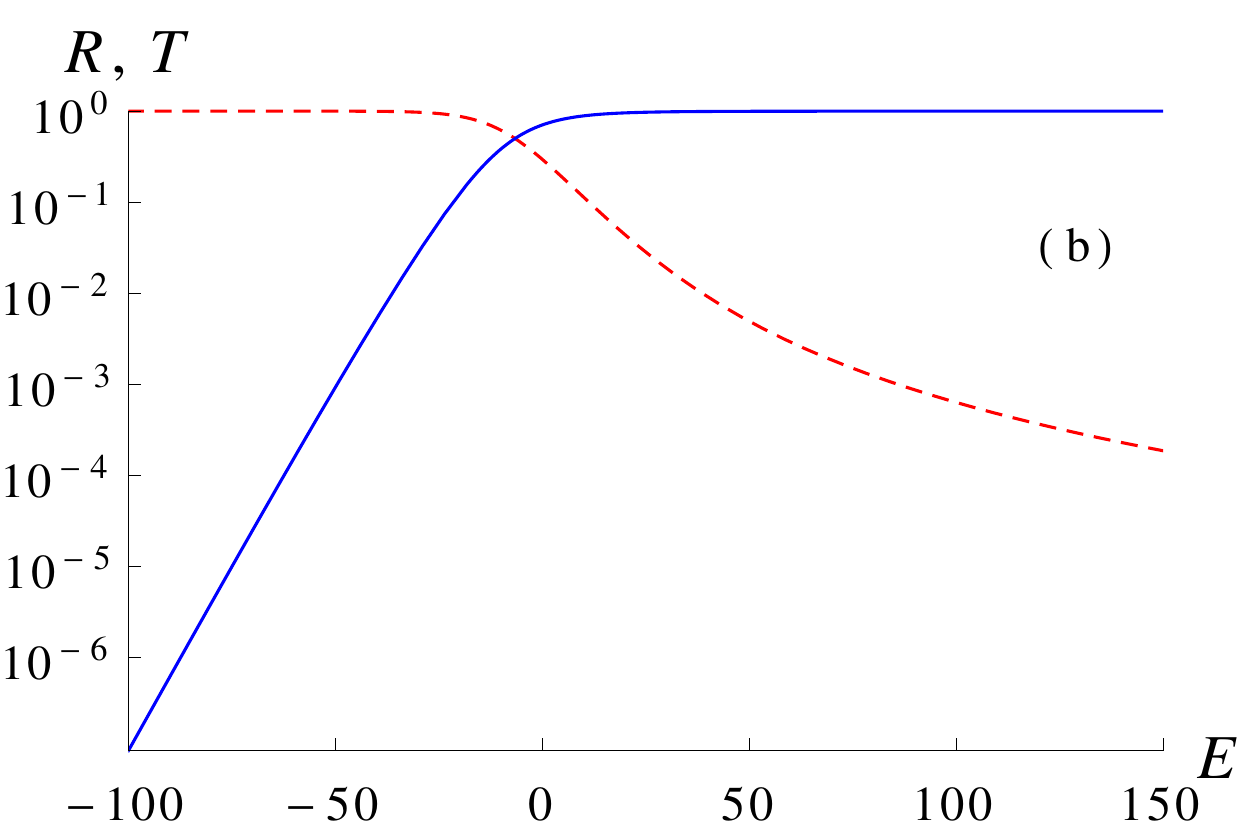}
	\hspace{0.5cm}
	\includegraphics[width=4.5cm,height=5 cm,scale=1.3]{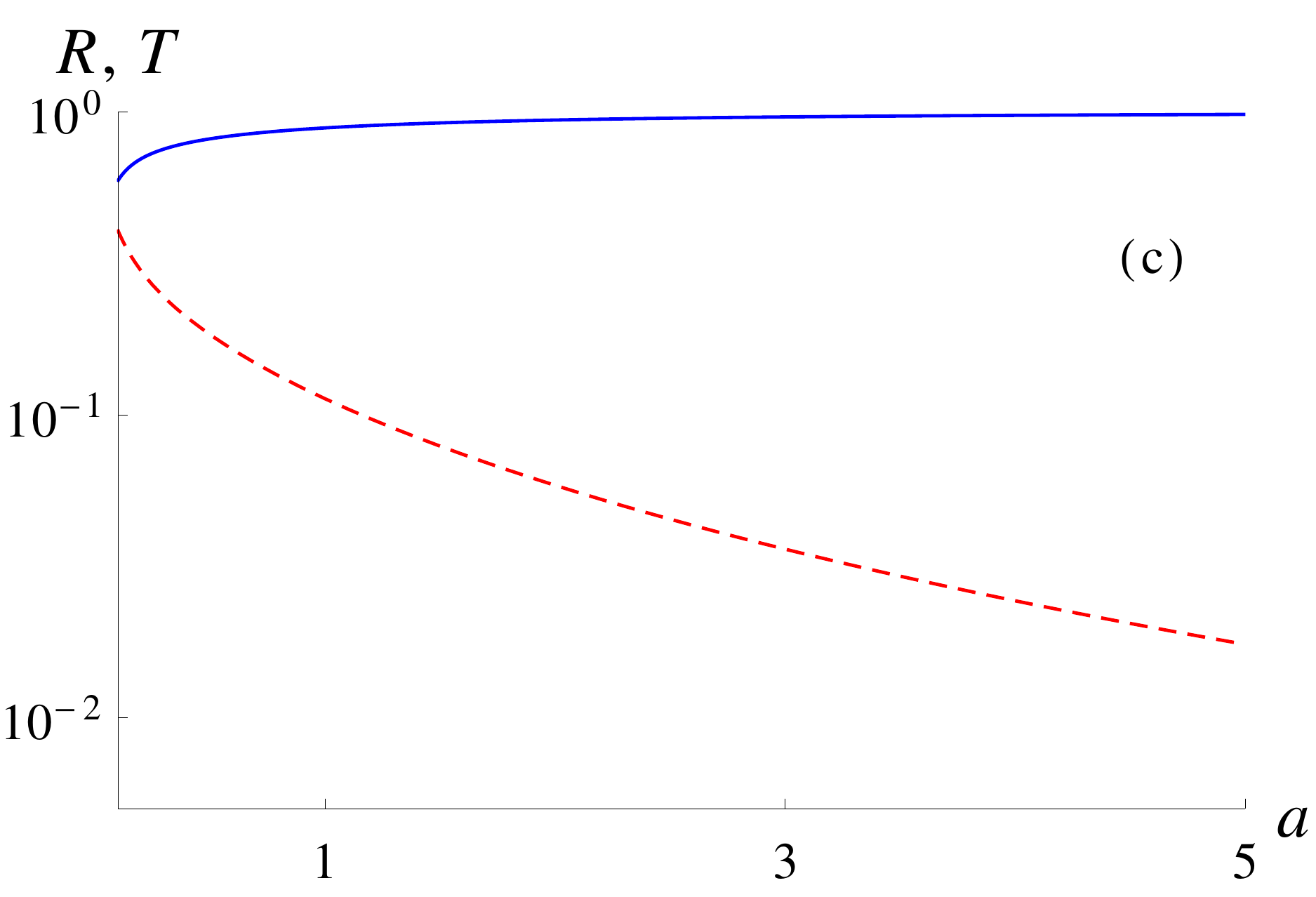}
	\caption{(a): The exponential bottomless potential barrier $V(x)=-V_0[\exp(2|x|/a)-1]$ (Eq. (1)) for $V_0=50$; $a=1$ (blue solid) and $a=5$ (red dashed) [scaled up by 500/3]. (b): Reflection $R(E)$ (red dashed) and the transmission $T(E)$ (blue solid) probabilities  for $V(x)$ in  Fig. 1(a) for $a=1$.(c): $R(E)$  and $T(E)$  for $E=10, V_0=50$ when  the parameter $a$ is varied. Notice that this potential does not become reflection/transmission less at a finite energy in any case.}
\end{figure}
In this paper, we study the possibility of bound states in the bottomless exponential barrier
( See Fig. 1(a)) 
\begin{equation}
V(x)=-V_0[e^{2|x|/a}-1], \quad V_0>0.
\end{equation}
This is so far the third bottomless potential which is exactly solvable. Its solutions are possible  in terms of well known cylindrical Hankel and Bessel functions, its reflection and transmission coefficients have been obtained and studied recently [27] (see Fig. 1(b,c)).  It is capable of raising some interesting issues which are un-noticed so far. Here, we show an existence of continuum of non-square-integrable degenerate states of definite parity in or above the exponential potential (1) for energies $E<V_0$. Next we show the surprising presence of discrete energy, square-integrable, definite-parity, non-degenerate states embedded in the said continuum of energies.

The Schr{\"o}dinger equation 
\begin{equation}
\frac{d^2\psi(x)}{dx^2}{+}[-\kappa^2+q^2 e^{2|x|/a}]\psi(x){=}0, ~~ \kappa{=}\frac{\sqrt{2m(V_0-E)}}{\hbar}, 
\end{equation}
and $q=\sqrt{2mV_0}/\hbar$, 
for this potential can be transformed to cylindrical Bessel equation [27] to get  two linearly independent solutions: 
\begin{equation}
\psi_+(x)=J_{\kappa a}(qae^{|x|/a}), \quad \psi_-(x)=J_{-\kappa a}(qae^{|x|/a}).
\end{equation}
These two solutions oscillate and vanish asymptotically. From these two solutions we construct two even and odd parity solutions such that $\psi'_e(0)=0$ and $\psi_o(0)=0$. These two definite parity degenerate  solutions for any energy in the continuum of energies $E<V_0$ when  $\kappa a \ne $ integer are
\begin{subequations}
\begin{align}
\psi_e(x)=C~[J'_{-\kappa a}(qa) J_{\kappa a}(z)-J'_{\kappa a}(qa) J_{-\kappa a}(z)], \\
\hspace{-1 cm}\psi_o(x){=}C\mbox{sign}(x)[J_{-\kappa a}(qa)J_{\kappa a}(z){-}J_{\kappa a}(qa)J_{-\kappa a}(z)],
\end{align}
\end{subequations}
where $z=qa e^{|x|/a}$.
Integer values of $\kappa a$ are the curious energies where the two linearly independent solutions (3) become linearly dependent, the states in (4) become identically zero. Only usual scattering states [27] exist at these energies, $R$ and $T$ remain finite see Figs. 1(b,c). It must be remarked that for energies $E>V_0$ all the solutions  of (1) are complex 
scattering states in terms of complex Hankel functions [27]  whose real and imaginary parts will be of definite-parity only at some special energies. But then real and imaginary parts of a complex function become ambiguous if an arbitrary complex (non-real) number $C$ multiplies it. For instance, let $\psi(x)=f(x)+ig(x)$ where $f(x)$ and $g(x)$ are of definite parities and let $C=a+ib$ where $a $ and $b$ are real and arbitrary. Then $\Re [\psi(x)]=f(x), \Im [\psi(x)]=g(x)$ have definite parities but $\Re [C\psi(x)]=af(x)-bg(x)$ and $\Im [C \psi(x)]=bf(x)+ag(x)$ will not have  definite parities.

 In Ref. [25], for $V(x)=-V_0 e^{2|x|/a}$, similar solutions as in (4) have been sought in order to bring out degenerated bound states  in one dimension. Here, we remark that by virtue of  the following integral [26]
\begin{multline}
\int_{s}^{\infty} \frac{J^2_{r} (t) dt}{t} = \frac{1}{2r}-4^{-r} s^{2r} \Gamma(2r)~ _2F_3[\{r,1/2{+}r\},\{1{+}r,1{+}r,1{+}2r\},-s^2], ~ \mbox{if}{~}r>0, 
\end{multline}
divergent otherwise; $\psi_+(x)$ is square-integrable but $\psi_-(x)$ is not so. Consequently, the degenerate states in (4) are not square-integrable and they  can form only a continuum of degenerate scattering states, as it should be. 

Next, we  look for the square integrable states in (1) by admitting the solution $\psi_+(x)$. Since the potential is  symmetric, we seek two separate definite parity solutions at two different energies as
\begin{subequations}
\begin{align}
\psi_e(x)= D J_{\kappa a} (qa e^{|x|}),~ \mbox{when}~ J'_{\kappa a}(qa)=0 ~\mbox{and}~\\ \psi_o(x)= D~ \mbox{sign}(x) J_{\kappa a} (qa e^{|x|}), ~\mbox{when}~ J_{\kappa a} (qa)=0.
\end{align}
\end{subequations}
We claim that these are square-integrable, definite-parity, non-degenerate states of (1) which
are embedded in the  energy continuum $E<V_0$. In Fig. 2, we plot $J'_{\kappa a}(qa)$ and $J_{\kappa a}(qa)$ (See Eq. (6)) as functions of energy to demonstrate their five number of zeros. This potential barrier therefore has five non-degenerate discrete energy states at $E=18.6108, 37.2630, 44.8253, 48.9214, 49.9988$ which are alternatingly of even and odd parities.
\begin{figure}[ht]
	\centering
	\includegraphics[width=14cm,height=7 cm]{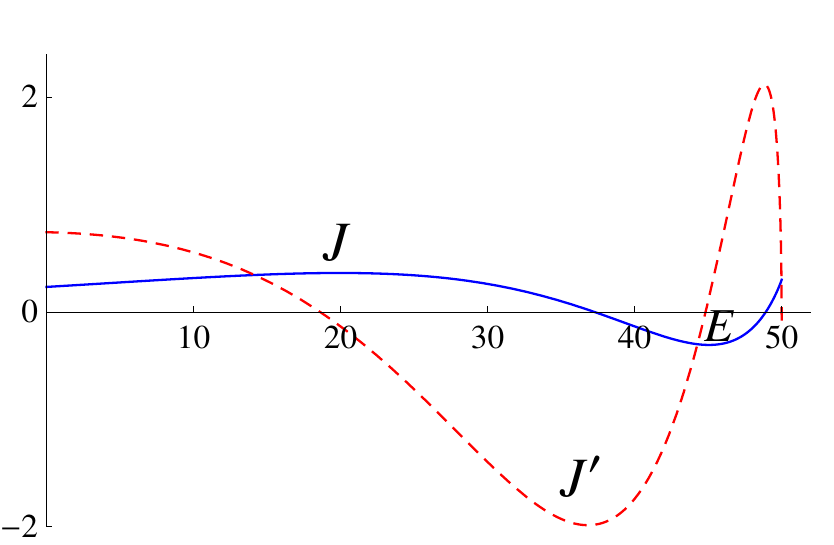}
	\caption{Using Eqs.~(6(a) and 6(b)), we plot $J_{\kappa a}(qa)$ (blue, solid line) and  $J'_{\kappa a}(qa)$ (red, dashed line) as functions of positive energy $(E{<}V_0)$.~Here, we take $2m/\hbar^2{=}1, V_0{=}50$ and $a{=}1$. Notice that there are overall five zeros of these  functions which are five eigenvalues of BIC above the barrier (1), and these are $E=18.6108, 37.2630, 44.8253, 48.9214, 49.9988$.}
\end{figure}
	
In Figs. 3(a) and 3(b), the first even and the first odd parity BICs  have been plotted for the case when $2m/\hbar^2=1, V_0=50$ and $a=1$. These states look curious bound states due to their numerous uncontrollable nodes. However, their knowledge could be worthwhile. It will be well to recall that even in 3D a BIC [1,2] has uncontrollable but fewer number of nodes.

\begin{figure}[]
	\centering
	\includegraphics[width=7cm,height=5 cm]{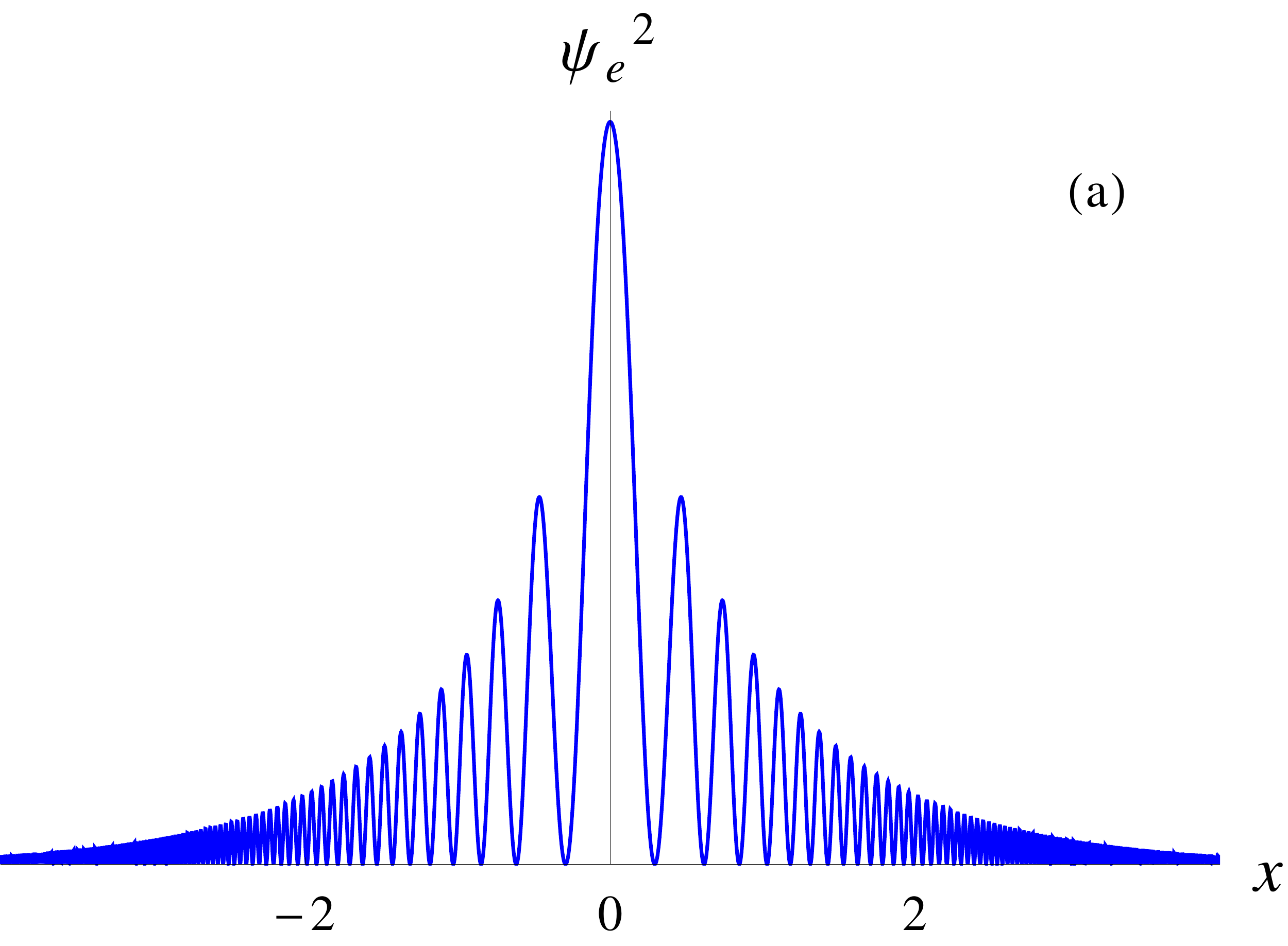}
	\includegraphics[width=7cm,height=5 cm]{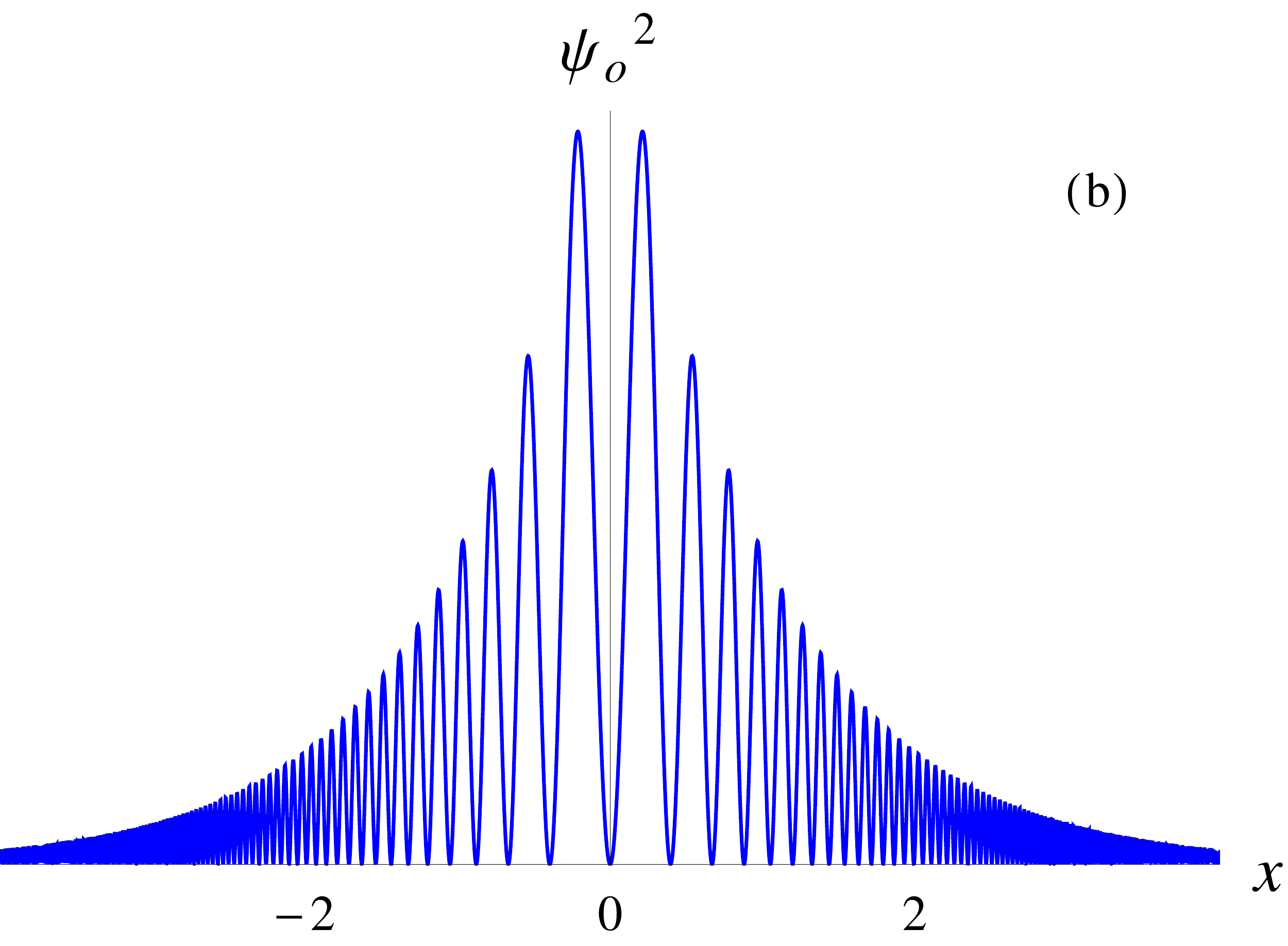}
	\caption{(a): Plot of the square of first even parity BIC  $\psi_e(x)$ (Eq. 4(a)) at $E=18.6108=E_*$ for the  barrier in Fig. 1.(a). (b): The square of first odd parity BIC $\psi_o(x)$ (Eq. (4b)) at $E=37.2630=E_*$. These states oscillates but vanishes at large asymptotic distances and it is square-integrable. At a nearby energy $E=E_*\pm \delta$ the state looses its definite parity,}
\end{figure}

A very interesting QES bottomless double barrier [24] has five real discrete energy eigenvalues known which lay in and above the well and the corresponding eigenstates are complex square-integrable scattering states such as $\psi(x)=\frac{\exp[(i\sinh x)/2]}{\sqrt{\cosh x}}$. 
The flux carried by them is non-zero, constant and energy-independent signifying a reflectionless state moving from one side to the other. The real and imaginary parts of these states are definite-parity degenerated bound states at one energy. 

Similarly, the first revelation [22] of the degenerated states in single or double bottomless barrier is also based on the reflectionless states such as $A\frac{\exp[i\gamma (x+x^3/3)]}{\sqrt{1+x^2}}$ and 
$\psi(x)=A\frac{\exp(i\gamma\sinh x)}{\sqrt{\cosh x}}.$ Therefore, the reflectionless states and the ensuing potential are connected [24] to the surprising existence of degeneracy in one dimension.

The structureless variation of $R(E)$ and $T(E)$ [27] as depicted in fig. 1(b,c) here rules out a possibility of transmission-zeros which have been observed in quasi-one dimensional structures [19]. The special integral values of $\kappa a=n$ when $E<V_0$ (see Eq. (4)) did appear curious in this regard when $E<V_0$, but we find that ordinary tunnelling occurs at these energies and $T(E)$ is non-zero and finite even at these energies.

In one dimension, we predict that simple bottomless potentials $V(x)=-|x|^{\nu}, \nu >2$ could easily be the models which will support square integrable, discrete energy, definite parity states above the barrier; They may be either degenerate or non-degenerate. In such cases, for asymptotic distances the energy being negligible the Schr{\"o}dinger equation becomes $\psi''+|x|^{\nu}\psi=0$. Its solutions are $\psi(x) =J_{1/(\nu+2)}[2(\nu+2)^{-1} |x|^{(\nu+2)}]$ whose asymptotic form works to $\psi(x) \sim \cos [\alpha |x|^{\nu+2} + \beta]/|x|^{\nu/4}$. Further, for the square-integrability of $\psi(x)$ we require $\nu >2$. But unfortunately even these simple models are not solvable analytically.

In this regard, the exponential barrier considered here like the parabolic and triangular barrier  has the usual feature of  not possessing reflectionlessness at any energy (see Fig. 1(b)) and hence it does not display degeneracy. The simple bottomless barrier $V(x)=-x^4$ [29] has been found to have real positive discrete energies where reflectivity of the barrier is zero. This potential is not analytically solvable and it would require special methods  to check if at those reflectivity zeros it
has degenerated bound states. Such a study is most welcome. In three dimensions, even a solitary  BIC has several nodes and the ensuing potential is oscillatory, but in one dimension the BIC has numerous nodes and the underlying potential is non-oscillatory. It may be remarked that the reported [22] degenerate bound states at $E=E_*$ in  QES potentials may also be suspiciously interesting as one may wonder if the same happens, as well, at a nearby energy $E=E_*\pm \delta$ leading to the formation of a continuum of such states. Lastly, we would like to re-emphasize that the bottomless exponential potential is the first exactly solvable model of BIC in one dimension. We hope that the present work will generate further interest in BIC in one dimension and the knowledge of such states could be at least worthwhile for a basic interest.

\section*{\Large{References}}


\begin{thebibliography}{40}
\bibitem{[16]} Wolfram Mathematica. Here $_2F_3$ is the regularized hypergeometric function.
\end{thebibliography}
\end{document}